\documentclass[%
 reprint,
 amsmath,amssymb,
 aps,
]{revtex4-1}
\pdfoutput=1
\usepackage{amsthm}
\usepackage{graphicx}
\usepackage{dcolumn}
\usepackage{bm}
\usepackage{extarrows}
\usepackage{float}
\usepackage[roman]{complexity}
\usepackage{color}
\usepackage{amstext}

\begin{document}

\title{Quantum mean centering for block-encoding-based quantum algorithm}

\author{Hai-Ling Liu $^{1,2}$}

\author{Chao-Hua Yu $^{3}$}

\author{Lin-Chun Wan $^{4}$}

\author{Su-Juan Qin $^{1}$}
\email{qsujuan@bupt.edu.cn}

\author{Fei Gao $^{1}$}
\email{gaof@bupt.edu.cn}

\author{Qiao-Yan Wen $^{1}$}

\affiliation{$^{1}$ State Key Laboratory of Networking and Switching Technology, Beijing University of Posts and Telecommunications, Beijing 100876, China}
\affiliation{$^{2}$ State Key Laboratory of Cryptology, P.O. Box 5159, Beijing, 100878, China}
\affiliation{$^{3}$ School of Information Management, Jiangxi University of Finance and Economics, Nanchang 330032, China}
\affiliation{$^{4}$ School of Computer and Information Science, Southwest University, Chongqing 400715, China}
\date{\today}

\begin{abstract}
Mean Centering (MC) is an important data preprocessing technique, which has a wide range of applications in data mining, machine learning, and multivariate statistical analysis. When the data set is large, this process will be time-consuming. In this paper, we propose an efficient quantum MC algorithm based on the block-encoding technique, which enables the existing quantum algorithms can get rid of the assumption that the original data set has been classically mean-centered. Specifically, we first adopt the strategy that MC can be achieved by multiplying by the centering matrix $C$, i.e., removing the row means, column means and row-column means of the original data matrix $X$ can be expressed as $XC$, $CX$ and $CXC$, respectively. This allows many classical problems involving MC, such as Principal Component Analysis (PCA), to directly solve the matrix algebra problems related to $XC$, $CX$ or $CXC$. Next, we can employ the block-encoding technique to realize MC. To achieve it, we first show how to construct the block-encoding of the centering matrix $C$, and then further obtain the block-encodings of $XC$, $CX$ and $CXC$. Finally, we describe one by one how to apply our MC algorithm to PCA and other algorithms.
\end{abstract}

\pacs{Valid PACS appear here}
\maketitle


\section{Introduction}
\label{Sec:introduction}

Quantum computing surpasses classical computing by taking advantage of the unique properties of superposition and entanglement in quantum mechanics, providing polynomial and even exponential speedups for specific problems, such as factoring integers~\cite{S1994}, unstructured database searching~\cite{LKG1997}, Hamiltonian simulation~\cite{BDGRS2007,BDAR2015,LGC2017,LGCI2019},
linear algebra~\cite{HHL,WY,LQ}, the subset sum problem~\cite{TSSP2022}, regression~\cite{NDS2012,W2017,YGW2019}, classification~\cite{RML2014,duan2017,TSVM}, clustering~\cite{LHJ2022}, correlation filter~\cite{KCF}.  Since its inception, quantum computing has become a booming research field. An introduction to more quantum algorithms can be seen in Refs.~\cite{MA2016,BJW2017}.

Mean Centering (MC) (also called zero centered or mean subtraction)~\cite{BS2003,HP2014} is the act of subtracting a variable's mean from all observations on that variable in the data set such that the variable's new mean is zero. It is an important data preprocessing technique, which has a wide range of applications in data mining, machine learning, and multivariate statistical analysis. For example, in Principal Component Analysis (PCA), to better describe the direction of the principal components~\cite{AHWL2010,BS2003}, it is necessary to perform eigen-decomposition on the mean-centered covariance matrix.

Driven by the progress in quantum computing, some quantum algorithms with significant acceleration compared with classical counterparts have been proposed to solve the classical problems involving MC, such as the well-known quantum PCA~\cite{SMP2014}, quantum Linear Discriminant Analysis (LDA)~\cite{IL2016}. However, these quantum algorithms assume that the original data set has been classically mean-centered. In fact, MC is not a simple process, and its complexity is  $O(n^{2})$, where $n$ comes from the size of the original data matrix $X\in \mathbb{R}^{n\times n}$. When the data set is large, this process will still be time-consuming. Therefore, it is meaningful to design an efficient quantum MC algorithm and study its application in PCA, LDA and other algorithms.

In Ref.~\cite{LLPS2022}, Li et al. adopted quantum inner product estimation algorithm~\cite{KK2019} and quantum
multiply-adder~\cite{zhou2017} to realize the MC process of the original data set and successfully applied it to Discriminative Canonical Correlation Analysis (DCCA) \cite{SCYS2008}. However, when Li et al.'s algorithm is applied to the classical problems involving MC, such as PCA, DCCA, these quantum algorithm relies on a strong assumption that at least half of the entries in the mean-centered data matrix have an absolute value greater than $m_{0}$, where $m_{0}>0$ is a constant. This limits the scope of application of the algorithm. In addition, Li et al.'s algorithm perform the quantum inner product estimation algorithm~\cite{KK2019} to compute the mean value of the original data set, which makes the quantum algorithm highly dependent on the inverse of the error $\epsilon$. For example, the dependence of quantum DCCA ~\cite{LLPS2022} is $1/\epsilon^{3}$. Therefore, an efficient quantum MC algorithm that does not require the above strong assumption and is more suitable for algorithms such as PCA, DCCA needs to be designed.

In this work, we propose an efficient quantum MC algorithm, which can get rid of the above strong assumption, and we successfully applied it to PCA, LDA, Canonical Correlation Analysis (CCA)~\cite{HSS2004}, DCCA, Ordinary Least Squares (OLS)~\cite{HGD2011}, etc. Specifically, we first adopt the strategy that the MC process of the original data can be achieved by multiplying by the centering matrix $C$~\cite{OB2021,BCWP2021}, that is, removing the row means, column means and row-column means of the original data matrix $X$ can be expressed as $XC$, $CX$ and $CXC$, respectively. This allows many classical problems involving MC, such as PCA, LDA, to directly solve the matrix algebra problems related to $XC$, $CX$ or $CXC$. Next, we can employ the block-encoding technique~\cite{LGCI2019,GSYLW2019,CSGJ2018} which has been successfully applied in quantum regression~\cite{CSGJ2018}, quantum classification~\cite{Shao2020} and other algorithms to realize MC. The core of our entire algorithm is to construct the block-encoding of the mean-centered data matrix. To achieve it, we first show how to construct the block-encoding of $C$, and then further obtain the block-encodings of $XC$, $CX$ and $CXC$ respectively. Finally, we describe one by one how to apply our quantum MC algorithm to PCA, LDA, and other algorithms. It is worth noting that our algorithm obtains the block encoding of the mean-centered data matrix, which makes our algorithm more suitable for algorithms involving matrix algebra problems, such as multidimensional scaling~\cite{Cox2008}, kernel machine learning~\cite{HTS2008,BCWP2021}, and extreme learning machine~\cite{CCCL2018}.

The remainder of the paper is organized as follows. In Sec.~\ref{Sec:Review}, we give a brief overview of the process of mean centering on the original data. In Sec.~\ref{Sec:Quantum}, we propose a quantum mean centering algorithm based on the block-encoding technique. In Sec.~\ref{Sec:Applications}, we present some applications of our quantum algorithm, such as PCA, LDA. Finally, we present our conclusion in Sec.~\ref{Sec:Conclusion}.

\section{Review of mean centering}
\label{Sec:Review}

Given an original data matrix $X=(\mathbf{x}_{1},\mathbf{x}_{2},\cdots,\mathbf{x}_{n})\in \mathbb{R}^{n\times n}$, it is shown as follows:
\begin{equation}
\begin{aligned}
 X&=\left[
\begin{array}{cccc}
x_{11} &x_{21} & \cdots & x_{n1} \\
x_{12} & x_{22} & \cdots & x_{n2} \\
\vdots & \vdots &\vdots & \vdots \\
x_{1n} & x_{2n}&\cdots & x_{nn}\\
 \end{array}
 \right],
 \end{aligned}
 \end{equation}
 then we can obtain the column means of $X$:
 \begin{equation}
 \begin{aligned}
 \mathbf{u}&=\left(\frac{x_{11}+\cdots+x_{1n}}{n},\cdots,\frac{x_{n1}+\cdots+x_{nn}}{n}\right)^{T}\\
 &\equiv\left(\bar{x}_{1\bullet},\cdots,\bar{x}_{n\bullet}\right)^{T}\in \mathbb{R}^{n},
 \end{aligned}
 \end{equation}
 and the row means of $X$:
 \begin{equation}
 \begin{aligned}
 \mathbf{v}&=\left(\frac{x_{11}+\cdots+x_{n1}}{n},\cdots,\frac{x_{1n}+\cdots+x_{nn}}{n}\right)^{T}\\
 &\equiv\left(\bar{x}_{\bullet1},\cdots,\bar{x}_{\bullet n}\right)^{T}\in \mathbb{R}^{n}.
 \end{aligned}
 \end{equation}

 Next, we can perform MC on $X$ to remove the column means, row means, and row-column means of $X$, respectively. The specific form can be expressed as follows:

$(1)$ Remove the column means of $X$: $x_{ij}-\bar{x}_{i\bullet}$;

$(2)$ Remove the row means of $X$: $x_{ij}-\bar{x}_{\bullet j}$;

$(3)$ Remove the row-column means of $X$: $x_{ij}-\bar{x}_{i\bullet}-\bar{x}_{\bullet j}+\bar{x}$,
where $\bar{x}=(x_{11}+\cdots+x_{1n}+\cdots+x_{n1}+\cdots+x_{nn})/n^{2}$ is the mean of all the entries of $X$, $i,j=1,\cdots,n$.

To simplify the MC process of $X$, scholars have found that it is achieved by projecting a matrix onto the column space orthogonal to the unit vector. Interestingly, this can be done by multiplying by the centering matrix
$C\in \mathbb{R}^{n\times n}$~\cite{OB2021,BCWP2021}, which is defined as follows:
\begin{equation}
\begin{aligned}
 C&=\left[
\begin{array}{ccccc}
1-\frac{1}{n} &-\frac{1}{n} & -\frac{1}{n}&\ddots  & -\frac{1}{n} \\
-\frac{1}{n} & 1-\frac{1}{n} & -\frac{1}{n} & \ddots & -\frac{1}{n} \\
\ddots & \ddots &\ddots & \ddots & \ddots\\
-\frac{1}{n} & -\frac{1}{n}&-\frac{1}{n} & \ddots & 1-\frac{1}{n}\\
 \end{array}
 \right]=I-\frac{1}{n}\mathbf{e}\cdot\mathbf{e}^{T},
 \end{aligned}
 \end{equation}
where $\mathbf{e}=(1,1,\cdots,1)^{T}\in \mathbb{R}^{n}$. It is worth noting that $C$ has many important properties, which can be seen in Refs.~\cite{OB2021,BCWP2021}.

 When $X$ is pre-multiplied by $C$, the column means of $X$ can be removed:
 \begin{equation}
\begin{aligned}
 CX&=\left[
\begin{array}{cccc}
x_{11}-\bar{x}_{1\bullet} &x_{21}-\bar{x}_{2\bullet}& \cdots & x_{n1}-\bar{x}_{n\bullet} \\
x_{12}-\bar{x}_{1\bullet} & x_{22}-\bar{x}_{2\bullet}& \cdots & x_{n2}-\bar{x}_{n\bullet}\\
\vdots & \vdots &\vdots & \vdots \\
x_{1n}-\bar{x}_{1\bullet} & x_{2n}-\bar{x}_{2\bullet}&\cdots & x_{nn}-\bar{x}_{n\bullet}\\
 \end{array}
 \right].
 \end{aligned}
 \end{equation}

When $X$ is post-multiplied by $C$, the row means of $X$ can be removed:
 \begin{equation}
\begin{aligned}
 XC&=\left[
\begin{array}{cccc}
x_{11}-\bar{x}_{\bullet1} &x_{21}-\bar{x}_{\bullet1}& \cdots & x_{n1}-\bar{x}_{\bullet1} \\
x_{12}-\bar{x}_{\bullet2} & x_{22}-\bar{x}_{\bullet2}& \cdots & x_{n2}-\bar{x}_{\bullet2}\\
\vdots & \vdots &\vdots & \vdots \\
x_{1n}-\bar{x}_{\bullet n} & x_{2n}-\bar{x}_{\bullet n}&\cdots & x_{nn}-\bar{x}_{\bullet n}\\
 \end{array}
 \right].
 \end{aligned}
 \end{equation}

 When $X$ are both pre-multiplied by $C$ and post-multiplied by $C$, the row-column means of $X$ can be removed:
\begin{equation}
\begin{aligned}
 &CXC=\\
 &\left[
\begin{array}{cccc}
x_{11}-\bar{x}_{1\bullet}-\bar{x}_{\bullet1}+\bar{x} & \cdots & x_{n1}-\bar{x}_{n\bullet}-\bar{x}_{\bullet n}+\bar{x}\\
x_{12}-\bar{x}_{1\bullet}-\bar{x}_{\bullet1}+\bar{x} & \cdots & x_{n2}-\bar{x}_{n\bullet}-\bar{x}_{\bullet n}+\bar{x}\\
\vdots & \vdots & \vdots \\
x_{1n}-\bar{x}_{1\bullet} -\bar{x}_{\bullet1}+\bar{x}&\cdots & x_{nn}-\bar{x}_{n\bullet}-\bar{x}_{\bullet n}+\bar{x}\\
 \end{array}
 \right].
\end{aligned}
\end{equation}

The complexity of the MC process of $X$ is $O(n^{2})$. This can be time-consuming when the original data set is large. And many classical algorithms, such as PCA, LDA, are related to the matrix algebra problems of $XC$, $CX$ or $CXC$. Therefore, it is meaningful to design an efficient algorithm to realize MC.

\section{A quantum mean centering algorithm based on block-encoding}
\label{Sec:Quantum}

In this section, we design an efficient quantum MC algorithm based on the block-encoding technique. The key point of this algorithm is to construct the block-encodings of $XC$, $CX$ and $XCX$, respectively. To achieve it, we first construct the block-encodings of $C$ and $X$ respectively, and further obtain the block-encodings of $XC$, $CX$ and $XCX$, respectively.

\subsection{Review of the block-encoding technique}
\label{Sec:block}

In this section, we review the block-encoding technique which can be found in Refs.~\cite{LGCI2019,GSYLW2019,CSGJ2018} for details.

$\mathbf{Definition}$ $\mathbf{1}$ (Block-encoding~\cite{LGCI2019}) Assume that $A$ is an $s$-qubits operator, $\alpha$, $\epsilon_{A}\in \mathbb{R}^{+}$, and $a\in \mathbb{N}$, then we say that the $(s+a)$-qubits unitary $U$ is an $(\alpha,a,\epsilon_{A})$ block-encoding of $A$ if it satisfies
\begin{equation}
\|A-\alpha(\langle0|^{\otimes a}\otimes I)U(|0\rangle^{\otimes a}\otimes I)\|\leq\epsilon_{A}.
\end{equation}
Note that $(1)$ when $A\in\mathbb{C}^{m\times n}$ where $n,m\leq 2^{s}$, we can construct an embedding matrix $A_{e}=
 \left[
\begin{array}{cc}
A& 0 \\
0 & 0\\
\end{array}
 \right]\in \mathbb{C}^{2^{s}\times 2^{s}}$; $(2)$ a unitary matrix is a $(1, 0, 0)$-block-encoding of itself, which we call
a trivial block-encoding.

Subsequently, Andr\'{a}s Gily\'{e}n et al.~\cite{GSYLW2019} showed how to implement the block-encoding of a linear combination of block-encoded operators and the product of block-encoded matrices respectively, as shown below:

$\mathbf{Definition}$ $\mathbf{2}$ (State preparation pair~\cite{GSYLW2019})
Let $\mathbf{y}\in\mathbb{C}^{m}$ and $\|\mathbf{y}\|_{1}\leq\beta$. The pair of unitaries $(P_{L},P_{R})$ is called an $(\beta,b,\varepsilon_{y})$-state-preparation-pair if $P_{L}|0\rangle^{\otimes b}=\sum_{j=1}^{2^{b}}c_{j}|j\rangle$ and
$P_{R}|0\rangle^{\otimes b}=\sum_{j=1}^{2^{b}}d_{j}|j\rangle$ such that $\sum_{j=1}^{m}|\beta(c_{j}^{\ast}d_{j})-y_{j}|\leq\varepsilon_{y}$ and for all $j\in m+1,\cdots,2^{b}$, we have $c_{j}^{\ast}d_{j}=0$.

$\mathbf{Lemma}$ $\mathbf{1}$ (Linear combination of block-encoded matrices~\cite{GSYLW2019})
Let $A=\sum_{j=1}^{m}y_{j}A_{j}$ be an $s$-qubits operator and $\varepsilon_{A}\in \mathbb{R}^{+}$. Assume that $(P_{L},P_{R})$ is an $(\beta,b,\varepsilon_{y})$-state-preparation-pair for $\mathbf{y}\in\mathbb{C}^{m}$, $W =\sum_{j=1}^{m}|j\rangle\langle j|\otimes U_{j}+((I-\sum_{j=1}^{m}|j\rangle\langle j|)\otimes I_{a}\otimes I_{s})$ is an $(s+a+b)$-qubits unitary such that for all $j=1,\cdots,m$, we have that $U_{j}$ is an $(\alpha,a,\varepsilon_{A})$-block-encoding of $A_{j}$. Then we can implement a $(\alpha\beta,a+b,\alpha\varepsilon_{y}+\alpha\beta\varepsilon_{A})$-block-encoding
of $A$, with a single use of $W,P_{R},$ and $P_{L}^{\dag}$.

$\mathbf{Lemma}$ $\mathbf{2}$ (Product of block-encoded matrices~\cite{GSYLW2019})
If $U$ is an $(\alpha; a;\delta)$-block-encoding of an $s$-qubit operator $A$ that can be
implemented in time $T_{1}$, and $V$ is an $(\beta; b;\varepsilon)$-block-encoding of an $s$-qubit operator $B$ that can be implemented in time $T_{2}$, then $(I_{b}\otimes U)(I_{a}\otimes V)$ is an $(\alpha\beta; a + b; \alpha\varepsilon+\beta\delta)$-block-encoding of $AB$ that can be implemented in time $O(T_{1}+ T_{2})$.

More information about the block-encoding technique can be found in Refs.~\cite{LGCI2019,GSYLW2019,CSGJ2018,TOTT2021}. This technique has been successfully applied to many algorithms, such as quantum linear system~\cite{LGCI2019,WCP2021}, quantum regression~\cite{CSGJ2018}, quantum classification~\cite{Shao2020}. In the following sections, we will introduce how to use the block-encoding technique to design quantum MC algorithm. For convenience, we define the base of the logarithm function as $2$, which can be abbreviated as $\log x$.

\subsection{A quantum mean centering algorithm based on the block-encoding technique}
\label{Sec:R}

The core of the whole quantum algorithm is to construct the block-encodings of $XC$, $CX$ and $XCX$, respectively. To achieve it, we first construct the block-encodings of $C$ and $X$ respectively.

\begin{figure}[!t]
\centering
	\includegraphics[width=8cm]{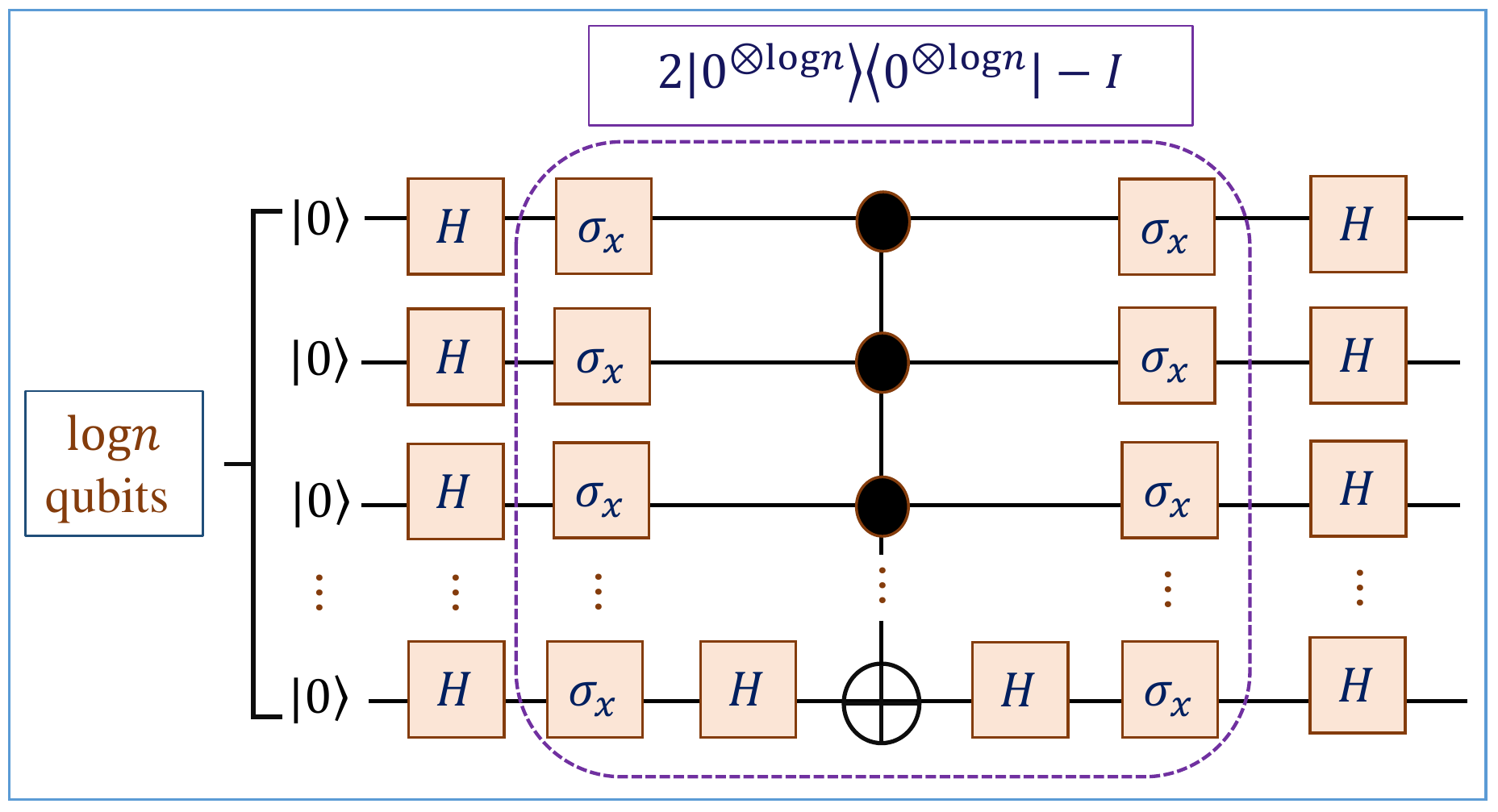}
	\caption{Quantum circuit for the unitary operator $U_{c}$. Here, $H$ and $\sigma_{x}$ denote Hadamard gate and Pauli operator $\sigma_{x}$, respectively.}
	\label{FIG:1}
\end{figure}

We find that $C$ can be express as
\begin{equation}
\begin{aligned}
C&=\frac{1}{2}\left(I-H^{\otimes \log n}(2|0^{\otimes \log n}\rangle\langle0^{\otimes \log n}|-I)H^{\otimes \log n}\right)\\
&\equiv\frac{1}{2}(I-U_{c}),
\end{aligned}
\end{equation}
where $H$ is the Hadamard gate and $2|0^{\otimes \log n}\rangle\langle0^{\otimes \log n}|-I$ is the conditional phase shift operator, which can be implemented using $O(\log n)$ quantum basic gates~\cite{NC2002}. The quantum circuit of $U_{c}$ is shown in Fig.~\ref{FIG:1}.

According to the definition of block-encoding, we can know that the unitary matrices $I$ and $U_{c}$ are an $(1,0,0)$-block-encoding of their own, which can be implemented with complexity $O(\log n)$. Thus, $C$ can be regarded as a linear combination of block-encoded operators.

Based on $\mathbf{Definition}$ $\mathbf{2}$, let $\mathbf{y_{c}}=(\frac{1}{2},-\frac{1}{2})\in \mathbb{R}^{2}$ and $\|\mathbf{y}_{c}\|_{1}=\beta_{c}=1$. And let $b_{c}=1, c_{j}=d_{j}=\sqrt{y_{c_{j}}}$, $j=1,2$, we have $P_{L}=P_{R}=SH$, i.e., $SH|0\rangle=\frac{1}{\sqrt{2}}(|0\rangle+i|1\rangle)$, where $S$ is a phase gate. Therefore, we can effectively construct an $(1,1,0)$-state-preparation-pair $(P_{L},P_{R})$ of $\mathbf{y_{c}}$ that satisfies the requirements of $\mathbf{Definition}$ $\mathbf{2}$. According to $\mathbf{Lemma}$ $\mathbf{1}$, we can construct an $(1,1,0)$-block-encoding of $C$ with complexity $O(\log n)$.

For the block-encoding of $X$, here we use the following lemma to achieve:

$\mathbf{Lemma}$ $\mathbf{3}$ (Implementing block-encoding from quantum data structure~\cite{LGCI2019,CSGJ2018})
If the data matrix $X\in \mathbb{R}^{n\times n}$ is stored in a quantum random access memory data structure~\cite{WLZZ2018}, then there exist two unitary operators $U_{\mathcal{M}}$ and $U_{\mathcal{N}}$ that can be implemented in time $O(\poly\log(n/\epsilon))$ such that $U_{\mathcal{M}}^{\dagger}U_{\mathcal{N}}$ is an $(\|X\|_{F},\poly\log n,\epsilon)$-block-encoding of $X$, where
\begin{equation}
\begin{aligned}
&U_{\mathcal{M}}:|i\rangle|0\rangle\mapsto|i\rangle|\mathbf{x}_{i}\rangle=|i\rangle\frac{1}{\|\mathbf{x}_{i}\|}\sum_{j=1}^{n}x_{ij}|j\rangle,\\
&U_{\mathcal{N}}:|0\rangle|j\rangle\mapsto\frac{1}{\|X\|_{F}}\sum_{i=1}^{n}\|\mathbf{x}_{i}\||i\rangle|j\rangle.
\end{aligned}
\end{equation}
Note that if $X\in \mathbb{C}^{n\times m}$ is also stored in the above data structure, then the unitary operator $U_{\mathcal{M}}^{\dagger}U_{\mathcal{N}}$ is also an $(\|X\|_{F},\lceil\log(m+n)\rceil,\epsilon)$-block-encoding of an extended Hermitian matrix $\bar{X}=X\otimes|0\rangle\langle1|+X^{\dagger}\otimes|1\rangle\langle0|$, which can be implemented in time $O(\poly\log(nm/\epsilon))$. A more detailed proof can be seen in $\mathbf{Lemma}$ $\mathbf{23}$ of Ref.~\cite{CSGJ2018}. In addition, if $X$ satisfies some specific conditions, such as $X$ is a sparse matrix or a purified density operator, we can also obtain the block-encoding of $X$~\cite{LGCI2019,GSYLW2019}.

According to the product of block-encoded matrices in $\mathbf{Lemma}$ $\mathbf{2}$, we can get an $(\|X\|_{F},\poly\log n,\epsilon)$-block-encodings of $XC$, $CX$, $CXC$ with complexity $O(\poly\log(n/\epsilon))$, respectively.

 Based on Refs.~\cite{LGCI2019,GSYLW2019,CSGJ2018}, we can solve the matrix algebra problems related to $XC$, $CX$ or $CXC$, such as matrix vector multiplication, and singular value estimation. Next, we will introduce the wide application of our quantum algorithm in PCA, LDA, and other algorithms.

\section{Applications of our quantum mean centering algorithms}
\label{Sec:Applications}

The classical problems involving MC, such PCA, LDA, CCA, DCCA and OLS, can be directly attributed to the matrix algebra problems related to $XC$, $CX$ or $CXC$. To solve matrix algebra problems efficiently, scholars have proposed a class of faster quantum algorithms based on the block-encoding framework~\cite{LGCI2019,GSYLW2019,CSGJ2018}. The core of this quantum algorithm is to construct the block-encoding of a specific matrix. Next, we will introduce one by one how to use our quantum MC algorithm to construct the block-encoding of matrices in PCA, LDA, CCA, DCCA and OLS.

\subsection{PCA}
\label{Sec:PCA}

PCA is the most common method for data dimension reduction, which can obtain low-dimensional data while maintaining the dispersion degree (variance) of original data as much as possible~\cite{AHWL2010}. To better describe the direction of the principal components~\cite{AHWL2010,BS2003}, PCA performs eigendecomposition on the mean-centered covariance matrix to capture effective low-dimensional data. The existing quantum PCA algorithm~\cite{SMP2014} assumes that $X$ has been classically mean-centered. Here, we start from $X$ to design a quantum PCA algorithm to avoid the above strong assumption. The optimization problem in PCA is expressed as:
 \begin{equation}
 \max_{U_{p}}Tr(U_{p}^{T}S_{t}U_{p}),s.t.,U_{p}^{T}U_{p}=I,
 \end{equation}
where
\begin{equation}
 S_{t}=\sum_{i=1}^{n}(\mathbf{x}_{i}-\mathbf{v})(\mathbf{x}_{i}-\mathbf{v})^{T}=XCX^{T}\in \mathbb{R}^{n\times n}
 \end{equation}
is the total scatter matrix (or called the mean-centered covariance matrix), $\mathbf{v}$ is the row mean of the
total data points, and $U_{p}=[\mathbf{u}_{1},\cdots,\mathbf{u}_{d}]\in \mathbb{R}^{n\times d} (1<d\ll n)$ is the the projection matrix whose columns $\{\mathbf{u}_{j}\}_{j=1}^{d}$ are the principal component directions of PCA.

The above optimization problem of PCA can be reduced to solve the eigenproblem for $S_{t}$. Then the eigenvectors corresponding to the first $1<d\ll n$ eigenvalues of $S_{t}$ are the principal component directions of PCA.

According to $\mathbf{Lemma}$ $\mathbf{2}$, we can obtain a unitary operator $Q$ is an $(\|X\|_{F}^{2}, poly\log n, 2\epsilon \|X\|_{F})$-block-encoding of $S_{t}$ that can be implemented in time $O(\poly\log(n/\epsilon))$. Then we can employ the optimal Hamiltonian simulation technique based on the block-encoding framework~\cite{LGCI2019,GSYLW2019,CSGJ2018} and quantum phase estimation algorithm~\cite{NC2002} to extract the eigeninformation of $S_{t}$, which takes $O([d\|X\|_{F}^{2}\poly\log(n/\epsilon)]/\epsilon)$ time complexity. Here, for convenience, let the error caused by using the phase estimation algorithm to estimate the eigenvalue is the same as the error caused by constructing the block-encoding of $X$, both of which are $\epsilon$.

\subsection{LDA}

LDA is a data dimensionality reduction algorithm involving data categories~\cite{BCNN2006}. LDA has been shown to be more effective than PCA in dealing with classification problems, i.e., classification with reduced dimensionality data~\cite{BHPK1997,CLZG1992}. The goal of LDA is to seek the optimal directions of projection on which the data points of different classes are far from each other while requiring data points of the same class to be close to each other. To achieve the above goal, the optimization problem in LDA can be expressed as follows:
 \begin{equation}
 \max_{U_{l}}Tr(U_{l}^{T}S_{b}U_{l}),s.t.,U_{l}^{T}S_{w}U_{l}=I,
\label{Eq:(13)}
 \end{equation}
where
\begin{equation}
 S_{b}=\sum_{k=1}^{c}n_{k}(\mathbf{v}_{k}-\mathbf{v})(\mathbf{v}_{k}-\mathbf{v})^{T}\in \mathbb{R}^{n\times n}
 \end{equation}
is the beween-class scatter matrix,
\begin{equation}
 S_{w}=\sum_{k=1}^{c}\sum_{i=1}^{n_{k}}(\mathbf{x}_{i}^{k}-\mathbf{v}_{k})(\mathbf{x}_{i}^{k}-\mathbf{v}_{k})^{T}\in \mathbb{R}^{n\times n}
 \end{equation}
is the within-class scatter matrix, where $n_{k}$ is the number of data points of the $k$th class, and $n=\sum_{k=1}^{c}n_{k}$ is the total number of data points, $\mathbf{v}_{k}$ is the row mean of the data points of the $k$th class, $\mathbf{v}$ is the row mean of the total data points, $\mathbf{x}_{i}^{k}$ is the $i$th data point of the $k$th class, $k=1,\cdots,c$, $i=1,\cdots,n$, and $U_{l}=[\vec{\mu}_{1},\cdots,\vec{\mu}_{d}]\in \mathbb{R}^{n\times d} (1<d\ll n)$ is the the projection matrix whose columns $\{\vec{\mu}_{j}\}_{j=1}^{d}$ are the optimal projection directions of LDA.

In particular, $S_{t}$ can be considered as the summation of $S_{b}$ and $S_{w}$, i.e., $S_{t}=S_{b}+S_{w}$ \cite{K1990}, therefore the optimization problem in Eq.~(\ref{Eq:(13)}) can be written as
 \begin{equation}
\max_{U_{l}}Tr(U_{l}^{T}S_{t}U_{l}),s.t.,U_{l}^{T}S_{w}U_{l}=I.
 \end{equation}
Hence, the optimal projection direction of LDA can be obtained by solving the generalized eigenvalue problem $(S_{t}; S_{w})$. Here, we adopt a quantum algorithm based on the block-encoding technology to solve the generalized eigenvalue problem of $(S_{t}; S_{w})$, as follows:

$\mathbf{Theorem}$ $\mathbf{1}$ (Solve the generalized eigenvalue problem of $(A;B)$ based on the block-encoding technology~\cite{shao2022})
Assume that $A$, $B$ are $n$-by-$n$ Hermitian matrices with $B$ positive definite. Let $U_{A}$
be an $(\alpha_{A}; q_{A}; \epsilon_{A})$-block-encoding of $A$ that is implemented in time $O(T_{A})$, and $U_{B}$ an $(\alpha_{B}; q_{B}; \epsilon_{B})$-block-encoding of $B$ that is implemented in time $O(T_{B})$. Denote $\kappa_{B}$ as the condition number of $B$,
the generalized eigen-pairs of $Ax = \lambda Bx$ as $\{(\lambda_{j},|E_{j}\rangle)|j=1,\cdots,n\}$. Given access
to copies of the state $\sum_{j}\beta_{j}|E_{j}\rangle$ that is prepared in time $O(T_{in})$, there is a quantum algorithm that
returns a state proportional to $\sum_{j}\beta_{j}|\tilde{\lambda}_{j}\rangle|E_{j}\rangle$, where $|\tilde{\lambda}_{j}-\lambda_{j}|\leq \epsilon$, in time
\begin{equation}
O\left(\kappa_{B}^{0.5}T_{in}+\frac{\alpha_{A}\kappa_{B}^{1.5}}{\|B\|\epsilon}(T_{A}+\frac{\alpha_{B}\kappa_{B}}{\|B\|})\right).
\end{equation}
 A detailed proof of this theorem can be seen in Ref.~\cite{shao2022}. Note that when $A$ and $B$ are non-Hermitian matrices, we can extend them to Hermitian matrices $\bar{A}=A\otimes|0\rangle\langle1|+A^{\dagger}\otimes|1\rangle\langle0|$ and $\bar{B}=B\otimes|0\rangle\langle1|+B^{\dagger}\otimes|1\rangle\langle0|$ respectively and when B is non-positive definite, we get its pseudo-inverse. Therefore, the above theorem can be extended to the case where $A$ and $B$ are not Hermitian matrices with $B$ not positive definite.

The key to solving the generalized eigenvalue problem of $(S_{t}; S_{w})$ by using $\mathbf{Theorem}$ $\mathbf{1}$ is to construct the block-encodings of $S_{t}$ and $S_{w}$, respectively. As far as we know, we have obtained the block-encoding of $S_{t}$. Next we will construct the block-encoding of $S_{w}$.

We find that $S_{w}$ can also be expressed as
 \begin{equation}
 S_{w}=\sum_{k=1}^{c}[X^{(k)}C^{(k)}(X^{(k)})^{T}],
 \end{equation}
 where $X^{(k)}=(\mathbf{x}_{k}^{(1)},\cdots,\mathbf{x}_{n_{k}}^{(k)})$ is the original data sub-matrix corresponding to each class, 
 $C^{(k)}=I_{n_{k}\times n_{k}}-\frac{1}{n_{k}}\mathbf{e}_{n_{k}}\mathbf{e}_{n_{k}}^{T}\in \mathbb{R}^{n_{k}\times n_{k}}$ is the centering matrix corresponding to each class, $k=1,2,\cdots,c$. Thus $S_{w}$ can be viewed as a linear combination of block-encoded operators.

Let $\mathbf{y}_{w}=(1,\cdots,1)\in \mathbb{R}^{c}$ and $\|\mathbf{y}_{w}\|_{1}=\beta_{w}=c$.
Let $b_{c}=\log c$, $c_{j}=d_{j}=1$, $j=1,\cdots,c$, we have $P_{L}=P_{R}=H^{\otimes \log c}$, i.e., $H^{\otimes \log c}|0^{\otimes\log c}\rangle=\frac{1}{\sqrt{c}}\sum_{j=1}^{c}|j\rangle$. Therefore, we can effectively construct an $(c,\log c,0)$-state-preparation-pair $(P_{L},P_{R})$ of $\mathbf{y}_{w}$ that satisfies the requirements of $\mathbf{Definition}$ $\mathbf{2}$. According to $\mathbf{Lemma}$ $\mathbf{1}$, we can construct an $(cf,\poly\log(c\tilde{n}),2c\epsilon f)$-block-encoding of $S_{w}$ with complexity $O(\poly\log(c\tilde{n}/\epsilon))$, where $f=\max_{k}(\|X^{(k)}\|_{F}^{2})$ and $\tilde{n}=\max_{k}(n_{k})$.

In short, we can solve the generalized eigenvalue problem $(S_{t}; S_{w})$. Compared with quantum LDA~\cite{IL2016}, our quantum algorithm does not require a strong assumption that the elements and the norms of the column vector of $S_{b}$ and $S_{w}$ can be accessed efficiently, respectively.

\subsection{CCA}

PCA and LDA are dimensionality reduction techniques for one set of variable data, while CCA is one of the unsupervised feature extraction techniques widely used in data pairs of two groups of variables, aiming at finding the optimal projection vectors pair to maximize the correlation between the two groups of variables~\cite{HSS2004}. Finding the optimal projection vectors pair in CCA can be transformed into solving the following generalized eigenvalue problem $(H_{x};H_{y})$:
\begin{equation}
\begin{aligned}
&\left[
\begin{array}{cc}
 & XCY^T \\
(XCY^T)^{T} & \\
 \end{array}
 \right]
 \left[
\begin{array}{cc}
 \mathbf{w}_{x} \\
  \mathbf{w}_{y}\\
 \end{array}
 \right]\\
 &=\lambda \left[
\begin{array}{cc}
 XCX^T & \\
 & YCY^T\\
 \end{array}
 \right]
 \left[
\begin{array}{cc}
 \mathbf{w}_{x} \\
  \mathbf{w}_{y}\\
 \end{array}
 \right]\\
 &\equiv H_{x}\mathbf{w}=\lambda H_{y}\mathbf{w},
 \end{aligned}
 \end{equation}
where $X\in \mathbb{R}^{n\times n}$ and $Y\in \mathbb{R}^{n\times n}$ are the original data matrices, $C\in \mathbb{R}^{n\times n}$ is the centering matrix, $\mathbf{w}=(\mathbf{w}_{x},\mathbf{w}_{y})^{T}\in \mathbb{R}^{2n}$ is the projection vectors pair.

According to $\mathbf{Theorem}$ $\mathbf{1}$, the core of solving the generalized eigenvalue problem $(H_{x};H_{y})$ is still to construct the block-encodings of $H_{x}$ and $H_{y}$ respectively.

For the block-encoding of $H_{x}$, we can first construct an $(\|X\|_{F}\|Y\|_{F}, \poly\log n, \epsilon(\|X\|_{F}+\|Y\|_{F}))$-block-encoding of $XCY^{T}$ with complexity $O(\poly\log(n/\epsilon))$, which is similar to $S_{t}$, where $\epsilon$ is generated by building the block-encodings of $X$ and $Y$. Then, according to $\mathbf{Proposition}$ $\mathbf{10}$ of Ref.~\cite{TOTT2021}, we also can obtain an $(\|X\|_{F}\|Y\|_{F}, \poly\log n, 2\epsilon(\|X\|_{F}+\|Y\|_{F}))$-block-encoding of $H_{x}$ with complexity $O(\poly\log(n/\epsilon))$~\cite{TOTT2021}.

For the block-encoding of $H_{y}$, we also obtain the unitary operator $P$ that is an $(\|Y\|_{F}^{2}, \poly\log n, 2\epsilon \|Y\|_{F})$-block-encoding of $YCY^{T}$, which requires time complexity $O(\poly\log(n/\epsilon))$. Furthermore, we can also generate the unitary operator $|0\rangle\langle0|\otimes Q+|1\rangle\langle1|\otimes P$ which is an $(\max\{\|X\|_{F}^{2},\|Y\|_{F}^{2}\},\poly\log n, 2\epsilon \min\{\|X\|_{F},\|Y\|_{F}\})$-block-encoding of $H_{y}$, where $Q$ is the block-encoding of $S_{t}=XCX^{T}$ and it takes $O(\poly\log(n/\epsilon))$ time complexity.

In a word, we can also design a quantum algorithm to solve CCA.

\subsection{DCCA}
\label{Sec:DCCA}

Although CCA can represent the information of two sets of variable data pairs to maximize the correlation between them, it ignores the difference between categories, which results in constraints on the performance of the algorithm. To solve the above problem, in 2008, Sun et al. proposed DCCA~\cite{SCYS2008}, aiming at finding the optimal projection vector pair to maximize the within-class correlations and minimize the between-class correlations by using the discriminant information of data points. Finding the optimal projection vector pair in DCCA can be transformed into solving the following generalized eigenvalue problems $(H_{d};H_{y})$:
\begin{equation}
\begin{aligned}
 &\left[
\begin{array}{cc}
 & XCECY^T \\
(XCECY^T)^{T} & \\
 \end{array}
 \right]
 \left[
\begin{array}{cc}
 \mathbf{v}_{x} \\
  \mathbf{v}_{y}\\
 \end{array}
 \right]\\
 &=\lambda \left[
\begin{array}{cc}
 XCX^T & \\
 & YCY^T\\
 \end{array}
 \right]
 \left[
\begin{array}{cc}
 \mathbf{v}_{x} \\
  \mathbf{v}_{y}\\
 \end{array}
 \right]\\
 &\equiv H_{d}\mathbf{v}=\lambda H_{y}\mathbf{v},
 \end{aligned}
 \end{equation}
 where $X\in \mathbb{R}^{n\times n}$ and $Y\in \mathbb{R}^{n\times n}$ are the original data matrices, $C\in \mathbb{R}^{n\times n}$ is the centering matrix, $E=diag(\mathbf{e}_{n_{1}}\mathbf{e}_{n_{1}}^{T},\cdots,\mathbf{e}_{n_{c}}\mathbf{e}_{n_{c}}^{T})\in \mathbb{R}^{n\times n}$ is the similarity matrix where $\mathbf{e}_{n_{k}}=(1,1,\cdots,1)^{T}\in \mathbb{R}^{n_{k}}$, $n_{k}$ is the number of data points of the $k$th class, and $n=\sum_{k=1}^{c}n_{k}$ is the total number of data points, and $\mathbf{v}=(\mathbf{v}_{x},\mathbf{v}_{y})^{T}\in \mathbb{R}^{2n}$ is the pairwise projection vectors of DCCA.

Similarly, the core of solving the generalized eigenvalue problem $(H_{d};H_{y})$ is still to construct the block-encodings  of $H_{d}$ and $H_{y}$ respectively. As far as we know, the block-encoding of $H_{y}$ has been constructed. Next, we will show how to construct the
block-encoding of $H_{d}$.

According to the form of $H_{d}$ and $\mathbf{Proposition}$ $\mathbf{10}$ of Ref.~\cite{TOTT2021}, we only need to construct the block-encoding of $XCECY^T$, and it can be regarded as the product of block-encoded operators. We have constructed the block-encodings of $X$, $C$, and $Y$, and then we will construct the block-encoding of $E$.

Since $E$ is a block diagonal matrix, we will construct the block-encodings of each diagonal block matrix $\mathbf{e}_{n_{k}}\mathbf{e}_{n_{k}}^{T}$, $k=1,\cdots, c$. It is worth noting that $\mathbf{e}_{n_{k}}\mathbf{e}_{n_{k}}^{T}=\sum_{j=1}^{n_{k}}V_{j}$, where the unitary operator $V_{j}=\sum_{j=1}^{n_{k}}|(k-j)\mod n_{k}\rangle|k\rangle$ is a $(1,0,0)$-block-encoding of itself, and the complexity is $O(\poly\log n)$. Similar to $S_{w}$, we can get the unitary operator $U_{e_{k}}$ that is an $(n_{k},\log n,0)$-block encoding of $\mathbf{e}_{n_{k}}\mathbf{e}_{n_{k}}^{T}$ with time $O(\poly\log n)$, $k=1,\cdots,c$. Then we can get that the unitary operator $\sum_{k=1}^{c}|k\rangle\langle k|\otimes U_{e_{k}}$ is an $(\tilde{n},\poly\log(cn),0)$-block encoding of $E$, and the time complexity is $O(\poly\log(c\tilde{n}))$, where $\tilde{n}= \max_{k}(n_{k})$.

According to $\mathbf{Lemma}$ $\mathbf{2}$, we can also produce an $(\tilde{n}\|X\|_{F}\|Y\|_{F},\log (c\tilde{n}),\epsilon\tilde{n}(\|X\| _{F}+\|Y\|_{F}))$-block-encoding of $XCECY^{T}$ with time complexity $O(\poly\log(cn/\epsilon))$. Similar to the method of $H_{x}$, we can obtain an 
$(\tilde{n}\|X\|_{F}\|Y\|_{F},\log (c\tilde{n}),2\epsilon\tilde{n}(\|X\|_{F}+\|Y\|_{ F}))$-block-encoding of $H_{d}$ with complexity $O(\poly\log(cn/\epsilon))$.

Similarly, based on $\mathbf{Theorem}$ $\mathbf{1}$, we also designed a quantum algorithm to solve DCCA. Compared with Li et al.'s algorithm~\cite{LLPS2022}, our algorithm does not need to call quantum inner product estimation algorithm~\cite{KK2019} twice to estimate the row mean of the original data set, which makes the inverse of the error dependence of our quantum DCCA algorithm drop from $1/\epsilon^{3}$ to $1/\epsilon$.

\subsection{OLS}

OLS is usually necessary to first perform mean centering on the original data to alleviate ``micro" multi-collinearity problem~\cite{ISPB2016,H2007}. The purpose of OLS is to find model parameters to minimize the sum of square deviations of the data in the model function~\cite{HGD2011}. Specifically, given the data points set $S=\{(\mathbf{x}_{i},y_{i})|\mathbf{x}_{i}\in \mathbb{R}^{n}, y_{i}\in \mathbb{R}\}_{i=1}^{n}$, the problem of OLS can be expressed as finding the parameters $\hat{\bm{\beta}}\in \mathbb{R}^{n}$ to minimize the following equation:
 \begin{equation}
\hat{\bm{\beta}}\equiv argmin_{\bm{\beta}}\|XC\bm{\beta}-\mathbf{y}\|^{2},
 \end{equation}
 where $X=(\mathbf{x}_{1},\cdots,\mathbf{x}_{n})\in \mathbb{R}^{n\times n}$ is the original data matrix, $C\in \mathbb{R}^{n\times n}$ is the centering matrix, $\mathbf{y}=(y_{1},\cdots,y_{n})^{T}\in \mathbb{R}^{n}$ is a vector. The above optimization problems has the following closed-form solution:
 \begin{equation}
\hat{\bm{\beta}}=(X^{T}CX)^{-1}X^{T}C\mathbf{y}.
 \end{equation}
Note that the above equation can be viewed as the product of the matrix $(X^{T}CX)^{-1}X^{T}C$ and the vector $\mathbf{y}$.

Combining with $\mathbf{Theorem}$ $\mathbf{34}$ in Ref.~\cite{CSGJ2018}, we only need to replace the weight matrix $W$ with $C$ to solve OLS. The complexity of quantum OLS algorithm is $\tilde{O}(\frac{\mu(A)\kappa_{A}}{\|A\|}\poly\log(n/\epsilon))$, where $A=\sqrt{C}X$, $\mu(A)=\|A\|_{F }$, $\kappa_{A}$ is the condition number of $A$. Similarly, our algorithm can be generalized to solve for weighted least squares and generalized least squares problems ~\cite{CSGJ2018}.

It is worth noting that our algorithm obtains the block-encoding of the mean-centered data matrix, which makes our algorithm more suitable for algorithms involving matrix algebra problems, such as multidimensional scaling~\cite{Cox2008}, kernel machine learning~\cite{HTS2008,BCWP2021}, and extreme learning machine~\cite{CCCL2018}.

\section{Conclusion}
\label{Sec:Conclusion}

In this paper, we presented a quantum MC for block-encoding-based quantum algorithm. In particular, we first constructed the block-encoding of $C$, and then further obtained the block-encodings of $XC$, $CX$ and $CXC$ respectively. Finally, our algorithm is successfully applied to PCA, LDA, CCA, DCCA and OLS, so that these corresponding quantum algorithms can get rid of the assumption that the original data matrix has been classically mean-centered. Compared with Li et al.'s algorithm~\cite{LLPS2022}, our algorithm can not only get rid of a strong assumption mentioned in Sec.~\ref{Sec:introduction}, but also reduce the dependence of the algorithm on the inverse of the error $\epsilon$. For example, in DCCA, our algorithm makes the error dependence of quantum DCCA drop from $1/\epsilon^{3}$ to $1/\epsilon$. It is worth noting that Li et al.'s algorithm produces a digital-encoded~\cite{pra2019} quantum state that can access the elements of the mean-centered data matrix in parallel. This may have advantages in other studies in the future.

We found an interesting phenomenon that MC can be achieved by calculating the mean value first and then subtracting the corresponding mean value from the original data, or by multiplying the centering matrix with the original data matrix. The complexity of both classical algorithms is $O(n^{2})$. Moreover, in many classical algorithms, such as PCA, the complexity of this process is often much smaller than the complexity of the algorithm itself, thus it is often ignored. However, for quantum algorithms, the strategy of multiplying the centering matrix with the original data matrix allows us to directly use the block-encoding technique to design the quantum MC algorithm. This makes this algorithm more suitable for  algorithms involving matrix algebra problems, such as multidimensional scaling~\cite{Cox2008}, kernel learning machine~\cite{HTS2008,BCWP2021}, extreme learning machine~\cite{CCCL2018}. This new phenomenon may shed some light on the work of quantum algorithms in other directions.

\begin{acknowledgments}
We thank Shanwei Ma and Changpeng Shao for useful discussions on the subject. This work is supported by the Beijing Natural Science Foundation (Grant No. 4222031), the National Natural Science Foundation of China (Grant Nos. 61976024, 61972048, 62006105), BUPT innovation and entrepreneurship support program (Grant No. 2021-YC-A206), and the Jiangxi Provincial Natural Science Foundation (Grant No.20202BABL212004).
\end{acknowledgments}

\end{document}